\documentclass[12pt]{article}
\usepackage{graphicx}
\usepackage{amsmath}
\usepackage{graphics}
\usepackage{epsfig}
\usepackage{amssymb}
\usepackage{color}
 \usepackage{setspace}

\usepackage[authoryear]{natbib}

\newcommand{\be}{\begin{eqnarray}}
\newcommand{\ee}{\end{eqnarray}}
\newcommand{\bea}{\begin{eqnarray*}}
\newcommand{\eea}{\end{eqnarray*}}

\newtheorem{theorem}{Theorem}[section]
\newtheorem{lemma}{Lemma}[section]

\newtheorem{corollary}{Corollary}[section]


\begin{document}

\title{Optimal designs for enzyme inhibition kinetic models}
\author{Kirsten Schorning, Holger Dette, Katrin Kettelhake, Tilman M\"oller \\
Fakult\"at f\"ur Mathematik \\
Ruhr-Universit\"at Bochum \\
44780 Bochum \\
Germany
}
\date{}
\maketitle

\begin{abstract}
In this paper we present a new method for determining optimal designs for enzyme inhibition kinetic models,
which are used to model the influence of the concentration of a substrate and an inhibition on the velocity of a reaction. The approach uses 
a nonlinear transformation of the vector of predictors such that  the model in the  new coordinates is given by an incomplete response surface model.
Although there exist no explicit solutions of the optimal design problem  for incomplete response surface models so far,
 the corresponding design problem in the new coordinates is substantially more transparent,  such that 
 explicit or numerical solutions  can be determined more easily.  The designs for the original problem 
 can finally  be found by an inverse transformation of the optimal designs determined for the  response surface model. 
We illustrate the method determining  explicit solutions for the $D$-optimal
design and for the optimal design problem for estimating the individual coefficients in a 
 non-competitive enzyme inhibition kinetic model.
\end{abstract}




\section{Introduction}
\label{sec1}
\def\theequation{1.\arabic{equation}}
\setcounter{equation}{0}

The Michaelis-Menten model
\begin{equation} \label{11}
\eta = \eta (S) =  \frac{V \cdot  S}{K_m+S}
\end{equation}
was introduced by  \cite{michment1913}  and is widely used
to represent   an enzyme kinetics reaction, where enzymes bind
substrates and turn them into products. Here $V$ and $S$
 denotes the maximum velocity of the reaction and the concentration  of
the substrate, respectively, while  $K_m$ is the value of $S$ at which half of the maximum
 velocity $V$ is reached, i.e., the Michaelis-Menten constant. Because of its importance
 the problem of designing experiments for  statistical analysis based on  the Michaelis-Menten model
 has found considerable interest in the literature and we refer to \cite{rasch1990,solwon1998,lopwong2002,detbie2003,matall2004}
 and \cite{dettkun2014}
 among many others, who investigate optimal and efficient
 designs  for the model \eqref{11} from different
 perspectives. 

The Michaelis-Menten model is very well justified in  the absence of enzyme inhibition
which are molecules that decrease the activity of enzymes.
However, many diseases require co-administration of  several
drugs and for this reason  new drugs are usually also
screened for their inhibitory potential.  Adequate modeling has to reflect this fact and therefore  the Michaelis-Menten model
has been extended to include the
effect of inhibitor concentration, say $I$. In these extended models $\eta$ is a function of the concentration of the substrate $S$ and the inhibition $I$, that is $\eta = \eta(S,I)$ and   one usually  distinguishes  between
competitive and non-competitive inhibition [see \cite{segel1993}, for example].
There does not exist much  literature on optimal designs for these type of models and we refer
to \cite{Youdim1019,bogackaetal2011,atkiboga2013} who mainly studied designs optimal with respect to determinant criteria, i.e., $D$- or $D_s$-optimal designs.
In most cases the optimal design problem is an extremely difficult optimisation problem and optimal designs have to be found numerically as the commonly used models  for competitive and non-competitive inhibition
are highly nonlinear and  therefore difficult to analyse mathematically. As a consequence not much - except of empirical findings - is known
about the structure  of optimal designs.

In this paper we present a new approach to derive optimal designs for regression models for  enzyme kinetics reactions involving inhibition which simplifies the analytical and numerical construction in this context substantially. 
The main idea is to use   a nonlinear transformation of the explanatory variables $\psi: (S,I) \to  (x,y) $ such that the model in the new variables
$x$ and $y$ is a simple (incomplete) multivariate polynomial  regression model (also called response surface model).
The  design problem can then be analysed in a linear response surface model
and in a second step the optimal designs from this model can  be transformed back into the $(S,I)$ coordinates
to obtain optimal designs for enzyme inhibition kinetic models.
We illustrate this approach determining optimal designs for a specific non-competitive enzyme inhibition kinetic model, but the general
idea is also applicable in other enzyme inhibition kinetic models as well. In particular, we provide a complete proof that a  design found in \cite{bogackaetal2011} is in fact $D$-optimal. Moreover, we  derive new optimal designs for
estimating the individual coefficients in this model. This optimisation problem is a substantially harder than the $D$-optimal
design problem and might be intractable in the classical parametrisation.
However, with the transformation introduced here  complete solution of the optimal design problem is possible.

Optimal designs for  multivariate quadratic regression  models have been widely studied in the literature and we refer to
the seminal papers of \cite{kono1962,farrell1967} and to \cite{draheipuk2000,drappuk2003} and \cite{detgri2014} for some more recent work among many others.
In these and many other references  optimal designs for the ``complete'' model are considered, which means the model under consideration
includes all multivariate polynomials  up to a given degree. In the simplest case of a cubic  polynomial in two variables, say $x$ and $y$, this means that the corresponding linear model
contains the functions
\begin{equation} \label{comp}
1,x,y,xy,x^2,y^2,xy^2,x^2y^2,x^3,y^3~.
\end{equation}
However, the  mapping $\psi$ introduced in this paper to transform design problems for enzyme inhibition kinetic models to designs problems for
linear response surface models usually results in ``incomplete" multivariate polynomial regression models,for example a cubic model involving only  the functions
\begin{equation} \label{incomp}
xy,xy^2,yx^2~.
 \end{equation}
 For such models the determination of optimal designs is substantially harder and there does not exist any solution so far. To our best knowledge
 - we  are only aware of two references,  where product designs for ``incomplete"  multivariate polynomials are determined [see \cite{detroe1996,detroe1997}].
 Although the  determination of optimal designs  (not restricted to a product structure) is extremely hard we are able to find solutions for the case \eqref{incomp}, which is the important one  for the  non-competitive enzyme inhibition kinetic model under consideration.

 The remaining part of this paper is organised as follows. In Section \ref{sec2} we introduce the regression model and some basic optimal
 design terminology. Section \ref{sec3} is devoted to the nonlinear transformation and its basic properties. Additionally we illustrate our approach determing $D$-optimal designs. In Section \ref{sec4} we use the new methodology to explicitly determine optimal designs for estimating the individual coefficients
 in a  non-competitive enzyme inhibition kinetic model. Finally, the  technical arguments  are deferred to an appendix in Section \ref{sec5}.

\section{A non-competitive model}
\label{sec2}
\def\theequation{2.\arabic{equation}}
\setcounter{equation}{0}

Let $S$ and $I$ denote the concentration of the substrate and inhibition, respectively, and let $\theta = (V,K_m, K_{ic})^T$ denote a vector of parameters. The non-competitive regression model considered in this paper is defined by
\be \label{model}
\eta(S, I, \theta) =  \frac{V \cdot  S}{(K_m+S)(1+\frac{I}{K_{ic}})}~,
\ee
where the explanatory variable $(S,I)$ varies in the  design space
$\mathcal{S} =[S_{\min}, S_{\max}] \times [I_{\min}, I_{\max}]$ defined by the the constants
$0 \leq S_{\min} < S_{\max}$, $0 \leq  I_{\min} < I_{\max}$. We assume that  observations
\be \label{model1}
Y_i =
\eta(S_i, I_i, \theta)  + \varepsilon_i ~,~i=1,\ldots ,n
\ee
are available at experimental conditions $(S_1, I_1),\ldots ,(S_n, I_n) \in {\cal S} $, where $\varepsilon_1 ,\ldots ,\varepsilon_n$ denote independent
normally distributed random variables with mean $0$ and variance $\sigma^2>0$.
Following \cite{kiefer1974} we define an approximate design
as probability measure, say $\xi$,  with finite support in $ {\cal S} $.
If the design $\xi$ has masses $\omega_i>0 $ at the different points $ (S_i, I_i) $ ($i =
1, \dots, k)$ and $n$ observations can be made by the experimenter,  the quantities
$\omega_i  n$ are rounded to integers, say $n_i$, satisfying $\sum^k_{i=1} n_i =n$, and
the experimenter takes $n_i$ observations at each location $(S_i, I_i)$  $ (i=1, \dots, k)$.
Now, under regularity assumptions, standard asymptotic theory shows that the maximum likelihood
estimator for the parameter $\theta$ is asymptotically unbiased with variance ${\sigma^2 \over n} M^{-1}(\xi,\theta) $,
where
\be \label{infmat}
M(\xi,\theta) = \int_{\mathcal{S}} \frac {\partial \eta (S,I,\theta)}{\partial \theta} \Big( \frac {\partial \eta (S,I,\theta)}{\partial \theta} \Big)^T d\xi (S,I).
\ee
denotes  the information matrix of a design $\xi$ in model \eqref{model} on the design space $\mathcal{S}$
and
\be \label{grad}
\frac {\partial \eta(S,I,\theta)}{\partial \theta}  = \frac {S}{(K_m + S)}\frac{1}{(1+I/K_{ic})} \Big (
1, - \frac {V}{K_m+S}, \frac {V \cdot I/K^2_{ic}}{1+I/K_{ic}} \Big)^T,
\ee
is the  gradient of the regression function with respect to the parameter $\theta$.
An optimal design maximises an appropriate functional, say $\Phi$, of the matrix $M(\xi,\theta)$ in the class of all designs
on the design space $\mathcal{S}$, which is called optimality criterion.
Several  criteria have been proposed in the literature to discriminate between competing designs [see for example \cite{pukelsheim2006}]
and  in  this paper we are interested in the widely used $D$-optimality criterion, which determines the
design such that
\be \label{eqdopt}
\Phi_D  \{ M(\xi,\theta) \}  =
\det \{ M(\xi,\theta) \}
\ee
 is maximal, and in optimal designs for estimating the individual coefficients, which maximise
\be \label{ind}
\Phi_{e_j}  \{ M(\xi,\theta) \}  = (e_j M^{-} (\xi,\theta)  e_j)^{-1} ~, ~~j=1,2,3~.
\ee
Here $e_j$ denotes the $j$th unit vector in $\mathbb{R}^3$, $M^{-} (\xi,\theta) $ is a generalised inverse of the matrix $M (\xi,\theta) $ and optimisation is only performed over the set
of designs, such that the information matrices  satisfy  the range inclusion $e_j \in $ Range($M (\xi,\theta) $). In this case the right-hand side of \eqref{ind} does not depend on the specific choice of the generalised inverse
[see \cite{pukelsheim2006}].
A design maximising \eqref{eqdopt} is called $D$-optimal design, while a design maximising \eqref{ind} is called $e_j$-optimal or optimal for
estimating the $j$th coefficient of the vector $\theta=(\theta_1,\theta_2,\theta_3)^T= (V,K_m, K_{ic})^T$ ($j=1,2,3$). For example, the $e_2$-optimal design is the
optimal design for estimating the Michaelis-Menten constant $\theta_2=K_m$.

Note that we  discuss  locally optimal designs, which require a-priori information about the unknown model parameters as  they
appear in  the  model in  a nonlinear  way [see \cite{chernoff1953}].  When preliminary knowledge regarding the unknown parameters of a nonlinear model is available, the application of locally optimal designs is well justified [see for example \cite{debrpepi2008}].  Locally optimal  designs  are also  typically  used  as benchmarks for commonly used
designs and often  serve as basis for constructing optimal  designs with respect to more sophisticated optimality criteria,  which are robust against a misspecification of the unknown parameters
[see \cite{pronwalt1985}  or \cite{chaver1995}, \cite{dette1997} among others]

\section{A nonlinear transformation and  $D$-optimal designs}
\label{sec3}
\def\theequation{3.\arabic{equation}}
\setcounter{equation}{0}

\subsection{ Transformation of the design space}
\label{sec31}
In this section we will provide a transformation   which makes the structure of the resulting design problem more visible. The transformation will be used to solve the $D$-optimal
design problem for the model \eqref{model} in Section \ref{sec32} and  for the solution of the $e_j$-optimal design problems in Section \ref{sec4}.
Recalling  the definition of the regression function in \eqref{model} we define a transformation by
\be \label{trafo2}
\left( \begin{array}{c}
 x \\ y
\end{array} \right)
= \psi(S,I) =
\left( \begin{array}{c}
\frac {S}{K_m+S} \\ \frac {1}{1+I/K_{ic}}
\end{array} \right) .
\ee
Note that \eqref{trafo2} defines a one-to-one mapping from the original design space
$\mathcal{S}=[S_{\min}, S_{\max}] \times [I_{\min}, I_{\max}]$ onto the rectangle
\be \label{xset}
\mathcal{X} =  [x_{\min}, x_{\max}] \times [y_{\min}, y_{\max}],
\ee
where the boundary points of the two intervals  are defined  by
\be \label{bp}
x_{\min} = \frac {S_{\min}}{K_m+S_{\min}}; \ x_{\max} = \frac {S_{\max}}{K_m+S_{\max}}; \
y_{\min} = \frac {1}{1+   I_{\max}/K_{ic}}; \ y_{\max} = \frac {1}{1+  I_{\min}/K_{ic}}. \
\ee
Moreover, the inverse transformation is given by
\be \label{invtrafo}
\begin{pmatrix}
S \\ I
\end{pmatrix}=
 \psi^{-1}(x, y) =
 \begin{pmatrix}
 \frac{xK_m}{1-x} \\
 \frac{K_{ic}(1-y)}{y}
 \end{pmatrix}.
\ee
A straightforward calculation shows that the gradient in \eqref{grad} can be represented by
$$
\frac {\partial \eta(S,I,\theta)}{\partial \theta}  = A(\theta) f(x,y),
$$
where the non-singular matrix $A(\theta)$ and the vector $f$ are given by
\be \label{Amat}
A(\theta) = \left(
\begin{array}{ccc}
1 & 0 & 0 \\
-\frac {V}{K_m} & \frac {V}{K_m} & 0 \nonumber \\
\frac {V}{K_{ic}} & 0 & - \frac {V}{K_{ic}}
\end{array}
\right),
\ee
\be \label{mat3}
f(x,y) = xy (1, x, y)^T,
\ee
respectively. If $\tilde \xi$ denotes the design on the design space $\mathcal{X}$ induced from the design $\xi$  by the transformation \eqref{trafo2}, then the information matrix $M(\xi,\theta)$ in \eqref{infmat} can be written as
\be \label{mtrafo}
M(\xi, \theta) = A(\theta) \tilde M(\tilde \xi) A^T(\theta),
\ee
where the matrix $\tilde M(\tilde \xi)$ is defined by
\be \label{mtilde}
\tilde M(\tilde \xi) = \int_{\mathcal{X}} f(x,y) f^T(x,y) d \tilde \xi (x,y).
\ee
As the transformation \eqref{trafo2} is one-to-one, any design $\xi$ on the design space $\mathcal{S}$ induces a unique design $\tilde \xi$ on $\mathcal{X}$ and vice versa.
Consequently, it follows from \eqref{mtrafo} that optimal designs maximising the functional $\Phi(M(\xi,\theta))$ can be obtained by  maximising the functional
$$
\Phi (A^T(\theta)\tilde M(\tilde \xi)A(\theta)).
$$

\subsection{$D$-optimal designs} \label{sec32}

$D$-optimal designs for the model \eqref{model} have been determined by \cite{bogackaetal2011}. In this section we explain how these designs can be obtained in a more transparent way from the regression model \eqref{mat3} using the transformation introduced in Section \ref{sec31}.
We will use the same method to derive new  optimal designs for estimating the individual coefficients in the subsequent Section \ref{sec4}.

In the case of $D$-optimality we have
$$
\Phi_D(M(\xi,\theta)) =\Phi_D(A^T (\theta) \tilde M(\tilde \xi) A(\theta)) =
(\det A(\theta))^2 (\det \tilde M(\tilde \xi))
$$
where the matrix $\tilde M(\tilde \xi)$ is given by \eqref{mtilde}.
Consequently, the $D$-optimal designs can be determined by first maximising the determinant $\det \tilde M(\tilde \xi)$ in the class of all designs $\tilde \xi$ on the design space $\mathcal{X}$ and then applying the inverse transformation \eqref{invtrafo}. Our first result provides the solution of this problem.

\begin{theorem} \label{lem1}
The $D$-optimal design maximising $\det \tilde M(\tilde \xi)$ in the class of all designs $\tilde \xi$ on the set $\mathcal{X}$ defined in \eqref{xset} is given by
\bea \label{dopt}
\tilde \xi^* = \left( \begin{array}{ccc}
\big(\max \{x_{\min}, \frac {x_{\max}}{2}\}, y_{\max}\big) & \big(x_{\max},\max  \{ \frac {y_{\max}}{2} , y_{\min} \}\big) &
(x_{\max}, y_{\max}) \\
\frac {1}{3} & \frac {1}{3} & \frac {1}{3}
\end{array}\right).
\eea
\end{theorem}

\medskip
\medskip

We can now use the inverse transformation \eqref{invtrafo} to obtain the $D$-optimal design on the original design space.

\begin{corollary} \label{rem1}
The locally $D$-optimal design for the non-competitive model \eqref{model} in the class of all designs $\xi$ on the set $\mathcal{S}$ has equal masses at the three points
$$\big( \max \{ S_{\min}, \frac {S_{\max}K_m}{S_{\max}+2 K_m} \}, I_{\min} \big)~;~
(S_{\max}, \min \{ K_{ic} + 2 I_{\min}, I_{\max}\})~;~ (S_{\max}, I_{\min}).
$$

\end{corollary}
\medskip

\textbf{Proof.} We use the inverse transformation \eqref{invtrafo} and  Theorem \ref{lem1} to obtain the $D$-optimal design on the original design space. For this purpose note that the first coordinate in \eqref{invtrafo} defines an increasing function of $x$, while the second coordinate is a decreasing function of $y$. Observing the relations in \eqref{bp} we obtain  that
$$
\psi^{-1} \big( \max \{ x_{\min}, \frac {x_{\max}}{2}\}, y_{\max}\big) =
\big( \max \{ S_{\min}, \frac {S_{\max}K_m}{S_{\max}+2K_m} \}, I_{\min} \big).
$$
\hfill $\Box$

\section{Optimal designs for estimating individual coefficients}
\label{sec4}
\def\theequation{4.\arabic{equation}}
\setcounter{equation}{0}

In this section we use the transformation \eqref{trafo2} to derive optimal designs for estimating the individual coefficients $V, K_m$ and $K_{ic}$ in the non-competitive regression model \eqref{model}. To our best knowledge all results presented here are new. \\
For this purpose consider the vector of parameters $\theta=(\theta_1, \theta_2, \theta_3)^T = (V, K_m,K_{ic})^T$ and
recall  that
an optimal design for estimating the coefficient $\theta_j$ maximises the expression \eqref{ind}
in the class of all designs $\xi$ satisfying the range inclusion
\be \label{rincl}
e_j \in \mbox{Range} (M(\xi,\theta)).
\ee
(here $e_j \in \mathbb{R}^3$ denotes the $j$th unit vector $(j=1,2,3)$).
Now recall the definition of the transformation in \eqref{mtrafo}. As the matrix $A(\theta)$ is non-singular, it follows that any generalised inverse $G$ of the matrix $M(\xi,\theta)$ induces a generalised inverse of the matrix $\tilde M(\tilde \xi)$ by the transformation $G \to A^T(\theta)GA(\theta)$ and vice versa. Similarly, if $\xi$ satisfies \eqref{rincl}, then $\tilde \xi$ satisfies the range inclusion
\be \label{rincl1}
c_j = B(\theta)e_j \in \mbox{Range} (\tilde M(\tilde \xi)),
\ee
 where
 \be \label{btilde}
B(\theta)  =(A(\theta))^{-1} ~=~
 \left(
\begin{array}{ccc}
1 & 0 & 0 \\
1 & \frac {K_m}{V} & 0 \\
1 & 0 & - \frac {K_{ic}}{V}
\end{array}
\right).
\ee
Consequently, an optimal design $\xi^*$ maximising the functional \eqref{ind} in the class of all designs $\xi$ satisfying \eqref{rincl} can be determined by maximising the functional
\bea \label{copt1}
(c^T_j \tilde M^- (\tilde{\xi}) c_j)^{-1}
\eea
in the class of all designs $\tilde\xi$ satisfying \eqref{rincl1} and transforming the design $\tilde \xi$ back onto the design space $\mathcal{S}$  using the inverse transformation \eqref{invtrafo}.\\
In the following two sections we derive the optimal designs for estimating the individual coefficients using this approach.
Section \ref{sec:mm_dis} contains the optimal designs for the Michaelis-Menten constant $K_m$ and the dissociation constant $K_{ic}$, that
is $j=2,3$. It will turn out that these two cases are almost the same in the transformed model \eqref{mat3}.
Section \ref{sec42} contains the optimal design for estimating the maximum velocity $V$  (that is $j=3$), where the determination of the optimal designs is much harder.

\subsection{Optimal designs for estimating the Michaelis-Menten constant and the dissociation constant}\label{sec:mm_dis}

The optimal design problem for estimating the Michaelis-Menten constant $K_m$ in the non-competitive model \eqref{model} corresponds to the choice $j=2$ in \eqref{ind}.  By the discussion at the beginning of this section we have to determine the design maximising the expression 
\be  \label{e2}
(c^T_2 \tilde M^-(\tilde \xi)c_2)^{-1}=  \big ( (B(\theta) e_2)^T \tilde M^-(\tilde \xi)(B(\theta) e_2) \big )^{-1}
=  \left(\frac {V}{K_m}\right)^2 ( e_2^T \tilde M^-(\tilde \xi) e_2)^{-1}
\ee
in the class of all designs satisfying the range inclusion \eqref{rincl1}  for $j=2$.
If an optimal design for estimating the parameter $K_{ic}$ is of particular interest, we use the criterion \eqref{ind} with $j=3$. Consequently, we have to maximise
\be \label{e3}
(c_3^T \tilde M^- (\tilde \xi)c_3)^{-1}  =\big ( (B(\theta)e_3 )^T \tilde M^- (\tilde \xi) B(\theta)e_3  \big)^{-1}=
\left(\frac {V}{K_{ic}}\right)^2 ( e^T_3 \tilde M^- (\tilde \xi)e_3 )^{-1}
\ee
in the class of all designs satisfying the range inclusion \eqref{rincl1}  for $j=3$.
\medskip
\begin{theorem}\label{thme2}
\begin{enumerate}
\item[(1)]  The optimal design maximising \eqref{e2} is of the form
\be \label{xitilde1}
\tilde \xi = \left( \begin{array} {cc}
(x_{\max},y_{\max})& (\overline{x} ,y_{\max}) \\
\frac {\overline{x} }{1+\overline{x} } & \frac {1}{1+\overline{x} }
 \end{array} \right)
 \ee
 where $\overline{x} = \max \big \{  x_{\min}, (\sqrt{2}-1)  x_{\max} \big \}$.
\item[(2)]  The optimal design maximising \eqref{e3} is of the form
 \be \label{xitilde3}
\tilde \xi = \begin{pmatrix}  (x_{\max},y_{\max})& (x_{\max}, \overline{y})  \\
\frac {\overline{y} }{1+\overline{y} } & \frac {1}{1+\overline{y} } \end{pmatrix}
\ee
 where $ \overline{y} = \max \big \{  y_{\min}, (\sqrt{2}-1)  y_{\max} \big \}$.
\end{enumerate}
\end{theorem}

\medskip

\begin{corollary} \label{cor2} ~~
\begin{enumerate}
\item[(1)] The optimal design for estimating the Michaelis-Menten constant in model \eqref{model} is given by
\be \label{xi}
\xi = \left( \begin{array} {cc}
 \big( S_{\max}, I_{\min}   \big)   &  \big( \overline{S} , I_{\min}  \big) \\
1- \omega & \omega
 \end{array} \right).
 \ee
 where
 \be
 \label{point}  \overline{S} &=& \max \Big\{ S_{\min}\  ,\  \frac{K_m S_{\max}  (\sqrt2 - 1) }{K_m +(2-\sqrt2)S_{\max} }  \Big\}   \\
\label{weight}   \omega &=&  \Big( 1 +  \max \Big\{ \frac{S_{\min}}{K_m + S_{\min}} \  ,\    \frac{  (\sqrt2 - 1) S_{\max}  }{K_m + S_{\max} }  \Big\}    \Big)^{-1}.
   \ee
\item[(2)]  The optimal design for estimating the parameter $K_{ic}$ in model \eqref{model} is given by
\be \label{xi3}
\xi =
\begin{pmatrix}
(S_{\max}, I_{\min}) & (S_{\max}, \overline{I}) \\
1- \omega & \omega
\end{pmatrix}
\ee
 where
 \bea
 \label{point3}  \overline{I} &=& \min \Big\{ I_{\max} ,I_{\min} (\sqrt{2}+1) + K_{ic}\sqrt{2}  \Big\}   \\
\label{weight3}   \omega &=&  \Big( 1 + K_{ic} \max \Big\{ \frac{1}{K_{ic} + I_{\max}}\  ,\    \frac{  (\sqrt2 - 1)}{K_{ic} + I_{min}} \Big\}    \Big)^{-1}.
 \eea
\end{enumerate}
\end{corollary}

{\bf Proof.}   We need to transform the design given by \eqref{xitilde1}  and  \eqref{xitilde3} to the actual design space $\mathcal{S}$.
For the design \eqref{xitilde1}  we note that  \eqref{trafo2} and \eqref{bp}  yield
 $$
 \overline x = \max \Big \{\frac{S_{\min}}{S_{\min} + K_m}, (\sqrt{2}-1) \frac {S_{\max}}{S_{\max}+K_m} \Big \},
 $$
 which gives for $\frac {1}{1+ \overline x}$ the weight in \eqref{weight}.
Due to the fact that the first component of the inverse $\psi^{-1}$ defined in \eqref{invtrafo} is increasing with respect to $x$ we obtain
 $$
 (\psi^{-1} (\overline x, y_{\max}) )_1= \overline S,
 $$
where $\overline S$ is defined in \eqref{point}.
Similar calculations for the other support point result in the
design $\xi$ in \eqref{xi}, which completes the   proof of the first part of Corollary \ref{cor2}. The second part
follows by similar arguments  and the details are omitted for the sake of brevity.
\hfill $\Box$
 \bigskip

\subsection{Optimal designs for estimating the maximal velocity}\label{sec42}

\begin{figure}[t]
\centering
 \includegraphics[width=0.6\textwidth]{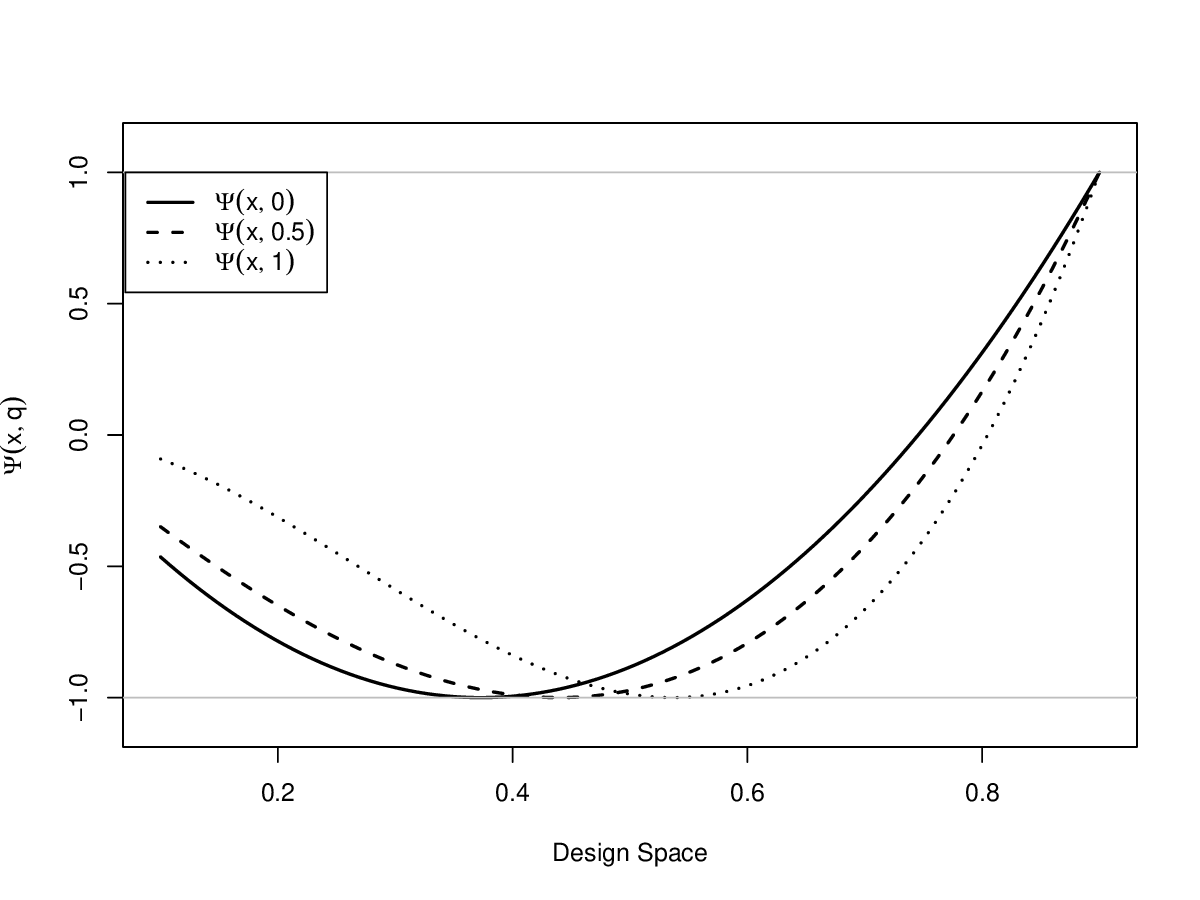}
\caption{\it The equi-oscillating polynomials $\Psi(x, q)$ for $q=0$ (solid line), $q=0.5$ (dashed line) and $q=1$ (dotted line) where the design space is given by $[0.1, 0.9]$. \label{fig1}}
\end{figure}

If the maximal velocity $V$ is the parameter of main interest, one can use the optimality criterion \eqref{ind} with $j=1$ to find the optimal design for a most precise estimation of $V$. Observing
\eqref{btilde}
we obtain from \eqref{rincl1} that $c_1 = (1,1,1)^T$
and the optimal design for estimating the maximum velocity can be found by maximising the criterion
\be \label{e1}
(c^T_1\tilde M^-(\tilde{\xi})c_1)^{-1}
\ee
in the class of all designs such that the range inclusion \eqref{rincl1} holds for $j=1$. The solution of this problem is substantially harder and
is based on the solution of an auxiliary optimal design problem for the  weighted polynomial regression model 
\begin{equation}\label{eq:helpeta}
\mathbb{E} [Y|x] =  x g(x, q)   (\theta_1 + \theta_2 x) ~,
\end{equation}
where  $q\in [0, 1]$ is a fixed but arbitrary constant, 
\be \label{eq:gfun}
g(x,q) =qx+1-q
\ee
 and  the design space is  given by  $ [x_{\min}, x_{\max}]$.
 Note  that $x_{\max} <1$, which follows from the definition \eqref{bp},  and  without loss of generality we assume $x_{\min} >0$ (see Section \ref{sec53} for further explanations).  
As the function $ g (x, q)$ is  positive on the interval $ [x_{\min}, x_{\max}]$ 
 it is easy to show that the set  $\{x g(x, q), x^2 g(x, q) \}$ is a Chebychev system on the  interval $[x_{\min}, x_{\max}]$ 
[see   \cite{karstu1966} for a definition of Chebyshev systems]. It then follows that there exists a unique function  (depending on the value of  $q$) of the form 
\begin{equation}
\Psi(x, q) = xg(x, q) (\psi_1 + \psi_2 x)
\end{equation}
(with constants $\psi_1, \psi_2 \in \mathbb{R}$) such that 
\begin{enumerate}
\item    $| \Psi(x, q)| \leq 1$ for all $x \in [x_{\min}, x_{\max}]$ . 
\item $\Psi(x_{\max}, q)=1$  and  there exists  exactly one additional  point $\bar{x}(q )\in [x_{\min}, x_{\max}] $ such that $\Psi(\bar{x}(q), q)=-1$. 
\end{enumerate}
This function $\Psi$ is called equi-oscillating polynomial in the literature and the points $x_{\max}$ and $\bar{x}(q )$  are called extremal points of $\Psi$. 
Some equi-oscillating polynomials  with their corresponding extremal points for different values of $q$ are depicted in Figure  \ref{fig1}. 
We obsere that $x_{\max}$ is always an extremal point and that the other extremal point increases with the vale of $q$. 
We are now in a position to describe the design maximising \eqref{e1} explicitly.

\begin{theorem} \label{theo:extrapc1}
~~\\
{\rm \bf {  a) }}
If   $x_{\max} \leq y_{\max}$, the optimal design maximising \eqref{e1} is of the form
\be \label{extrapol1}
\tilde{\xi}(q^*) =   \left( \begin{array} {cc}
 \big(\bar{x}, g(\bar{x},q^*)\big)   &  \big(x_{\max} ,y_{\max}   \big) \\
\omega(q^*)  &1-  \omega(q^*) 
 \end{array} \right)
 \ee
where
\begin{equation}\label{eq:helpweight}
\omega(q) = \frac{x_{\max}g(x_{\max},q) (1-x_{\max}) }{x_{\max}g(x_{\max},q) (1-x_{\max}) + \bar{x}(q) g(\bar{x}(q), q)(1-\bar{x}(q))}
\end{equation}
and  
\be \label{eq:qstar}
q^*   =  \frac{1-y_{\max}}{1-x_{\max}} . 
\ee
{\rm \bf {(b) }} If $x_{\max} \geq y_{\max}$, the optimal design maximising \eqref{e1}  can be obtained from part (a)  by interchanging the roles of $x$ and $y$.
\end{theorem}

\bigskip

\begin{figure}[t]
\centering
 \includegraphics[width=0.4\textwidth]{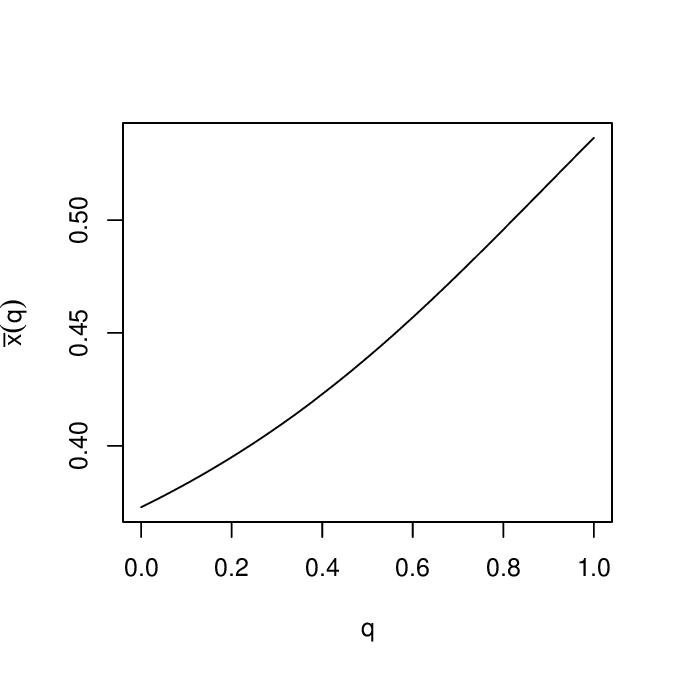}
  \includegraphics[width=0.4\textwidth]{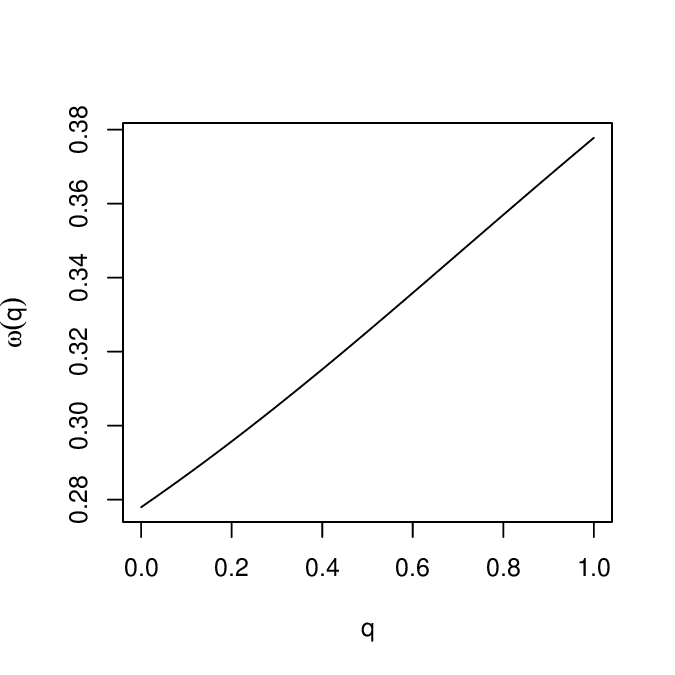}
\caption{The support point $\bar{x}(q)$ (left panel) and the corresponding weight $\omega(q)$ (right panel) of the $c$-optimal design given by \eqref{extrapol1} in dependence of the value $q$ for $x_{\max}=0.9<y_{\max}$. \label{fig2}}
\end{figure}

In Figure \ref{fig2} we display the support point $\bar{x}(q)$ and the corresponding weight $\omega(q)$ of the $c$-optimal design in \eqref{extrapol1} 
in dependence of the value $q$ for $x_{\max}=0.9<y_{\max}$. We observe that the support point and corresponding weight are increasing with the value of $q$.

Using similar arguments as given in Section \ref{sec3} the optimal design for estimating the maximum velocity can be obtained by 
the inverse transformation \eqref{invtrafo} from  the design in Theorem \ref{theo:extrapc1}. The explicit formulas in the general 
case are omitted for the sake of brevity and we restrict ourselves to the case  $I_{\min}=0$, which is most important from a practical
 point of view.

 \begin{corollary} If $I_{\min}=0$ the optimal design for estimating the maximum velocity in the non-competitive 
 regression model \eqref{model} is given by 
\begin{equation*}
\xi ^*= \begin{pmatrix}  (\bar{S}, I_{\min})  & (S_{\max}, I_{\min}) \\
\omega & 1-\omega \end{pmatrix},
\end{equation*}
where
\begin{eqnarray*}
\overline{S} &=& \max \Big\{ S_{\min}\  ,\  \frac{K_m S_{\max}  (\sqrt2 - 1) }{K_m +(2-\sqrt2)S_{\max} }  \Big\}   \\
\label{weight2}   \omega &=&\frac{S_{\max}(K_m +\overline{S})^2}{S_{\max}(K_m +\overline{S})^2 + \overline{S} (K_m +S_{\max})^2}.
   \end{eqnarray*}
\end{corollary}

\textbf{Proof.} If $I_{\min}=0$ it follows from  \eqref{trafo2} that  $y_{\max}=1$, which gives $q=0$ and for the function $g$ in \eqref{eq:gfun}  $g(x,0) \equiv 1$.
The equi-oscillating polynomial  corresponding to the Chebyshev system $\{ x,x^2 \} $ on the interval  $[x_{\min}, x_{\max}] $ is given 
by
\begin{equation*}
\underline{\Psi} (x,0) = \begin{cases}   -\big (\frac{x}{\bar{x}}\big)^2 + 2\frac{x}{\bar{x}}  & \mbox{ if }   x_{\min} \leq x_{\max}(\sqrt{2}-1)  \\
\frac{(x_{\min}+ x_{\max})}{x_{\min}x_{\max}(x_{\max}-x_{\min})} x^2 -\frac{(x^2_{\min}+ x^2_{\max})}{x_{\min}x_{\max}(x_{\max}-x_{\min})} x & \mbox{ if } x_{\min} >  x_{\max}(\sqrt{2}-1)
\end{cases}
\end{equation*}
where $\bar{x}= \bar{x}(0)= x_{\max}(\sqrt{2}-1)$. Here the equi-oscillating polynomial in the case  $x_{\min} \leq x_{\max}(\sqrt{2}-1)  $ 
is determined  in the proof of Theorem  \ref{theo:extrapc1}  (see Section \ref{sec53})
and the remaining case follows by similar arguments.
Consequently,   Theorem \ref{theo:extrapc1} shows that the 
 the  design maximising \eqref{e3} is given by
\begin{equation}\label{eq:ymax=1}
\tilde\xi= \begin{pmatrix}  \big (\max(x_{\min}, \bar{x}), y_{\max} \big)  & (x_{\max}, y_{\max}) \\
\frac{x_{\max}(1-x_{\max})}{x_{\max}(1-x_{\max}) + \bar{x} (1-\bar{x})} &  \frac{ \bar{x} (1-\bar{x})}{x_{\max}(1-x_{\max}) + \bar{x} (1-\bar{x}) } \end{pmatrix},
\end{equation}
and the assertion now follows by the inverse  transformation \eqref{invtrafo}. \hfill $\Box$

\bigskip
\bigskip

{\bf Acknowledgements} The authors would like to thank Martina
Stein, who typed parts of this manuscript with considerable
technical expertise. This work has also been supported in part by the
Collaborative Research Center ``Statistical modeling of nonlinear
dynamic processes'' (SFB 823, Teilprojekt C2) of the German Research Foundation
(DFG).

\setstretch{1.15}
\setlength{\bibsep}{1pt}
\begin{small}
 \bibliographystyle{apalike}
\itemsep=0.5pt
 \bibliography{paper_tilman_lit}
\end{small}

\section{Appendix: Proofs} \label{sec5}
\subsection{Proof of Theorem \ref{lem1}}\label{sec51}
A further transformation shows that
\bea 
f(x,y) = xy(1,x,y)^T = C f \Big( \frac {x}{x_{\max}}, \frac {y}{y_{\max}} \Big),
\eea
where the matrix $C$ is given by the diagonal $3 \times 3$ matrix
$
C= x_{\max} y_{\max} \cdot \mbox{diag} (1, x_{\max}, y_{\max}).
$
Consequently, we may assume without loss of generality
\bea 
0 \leq x_{\min} < x_{\max} = 1; \qquad 0 < y_{\min} < y_{\max} = 1.
\eea
For the sake of brevity we now restrict the discussion to the case that $0\leq x_{\min}, y_{\min} < \tfrac{1}{2}$, the other cases can be treated similarly.
In this case the resulting design
\begin{equation}\label{doptspez}
\tilde\xi^*= \begin{pmatrix}(\tfrac{1}{2},1) & (1,\tfrac{1}{2}) & (1, 1) \\ \tfrac{1}{3} & \tfrac{1}{3} & \tfrac{1}{3}, \end{pmatrix}
\end{equation}
and, by the equivalence theorem for $D$-optimality this design is $D$-optimal if and only if the inequality
\begin{eqnarray} \nonumber
\kappa(x, y) &=&  x^2y^2 \bigl( 1, x, y\bigr)\tilde{M}^{-1}(\tilde\xi^*)  \bigl( 1, x, y\bigr)^T -3  \\
 &=& 3x^2 y^2 (20 x^2  -44x + 8 xy + 20 y^2 - 44 y  +41) -3 \leq 0
\label{Dinq}
\end{eqnarray}
holds for all $(x, y)\in \mathcal{X} =  [x_{\min}, 1]\times[y_{\min}, 1]$.
Note that the function $\kappa$ is symmetric in $x$ and $y$.
We now calculate the points of $\kappa$ in $\mathcal{X}$ where the maxima are attained and show that the inequality \eqref{Dinq} holds for these points. \\
The stationary points of $\kappa$ in $(x_{\min}, 1) \times (y_{\min}, 1)$  are given by
\begin{equation*}
(x_1, y_1) = (\tfrac{1}{172}(55 - \sqrt{73}), \tfrac{1}{172}(55 - \sqrt{73})), \quad
(x_2, y_2) = (\tfrac{1}{172}(55 + \sqrt{73}), \tfrac{1}{172}(55 + \sqrt{73})).
\end{equation*}
Considering the Hessian matrix for these points it follows that $(x_1, y_1)$ is a saddle point of $\kappa$ whereas $(x_2, y_2)$ is a minimum. Consequently, the maxima of the function $\kappa$ are attained at the boundary of $\mathcal{X}$. \\
Because of the symmetry of $\kappa$ it is sufficient to calculate the maxima of the functions
\be
\kappa_1(x) &=& \kappa(x, y_{\min})  \label{kap1} = 60y_{\min}^2\ x^4  + (24y^3_{\min} -132y^2_{\min} )\ x^3  \\
 &&~~~~~~~~~~~~+~ (60 y_{\min}^4  - 132 y^3_{\min}  +123 y^2_{\min})\ x^2 -3. \nonumber  \\
\kappa_2(x) &=& \kappa(x, 1) = 60 x^4 -36 x^3 + 17 x^2 - 3  \label{kap2}
\ee
 for $x \in [x_{\min}, 1]$. We start with the maximisation of $\kappa_1(x)$.
 For $0<y_{\min}\leq \tfrac{1}{2}$ the only local extremum of $\kappa_1$ is a minimum attained in $x^*=0$ with value $\kappa_1(0)= -3$. Consequently, $\kappa_1$ is increasing for $x> 0$ and its maximum on $[x_{\min}, 1]$ is attained in $x^*=1$.   \\
The stationary points of the function $\kappa_2$ in \eqref{kap2} are given by
$
x= 0, \ x= \tfrac{1}{2}, \ x = \tfrac{17}{20}.
$
Since $\kappa_2$ is a polynomial of degree $4$ with positive leading coefficient the locally minima of $\kappa_2$ are attained at  $x=0$ and $x= \tfrac{17}{20}$. Consequently, the maxima of $\kappa_2$ on $[x_{\min}, 1]$ are attained in $x^* =   \tfrac{1}{2}$ and $x^* = 1$.
Using the symmetry the maxima of the function $\kappa$ given in \eqref{Dinq} on $\mathcal{X}$ are given by
\begin{equation*}
(x^*_1, y^*_1) = (1, y_{\min}), \quad (x^*_2,  y^*_2) = (\tfrac{1}{2}, 1), \quad (x^*_3, y^*_3) = ( 1, \tfrac{1}{2}), \quad (x^*_4, y^*_4) = (1, 1).
\end{equation*}
Obviously,  $  \kappa(x^*_2,  y^*_2) =\kappa(x^*_3,  y^*_3) = \kappa(x^*_4,  y^*_4)= 0$ and additionally
  $\kappa(x^*_1, y^*_1) \leq  0 $ for $y_{\min} \leq \tfrac{1}{2}$, since the function
$$h(y_{\min}) = \kappa_1(1) = 60y^4_{\min} -108y^3_{\min} + 51y^2_{\min} - 3$$
is increasing on $[0, \tfrac{1}{2}]$ with $h(\tfrac{1}{2}) = 0$.
Consequently, the inequality \eqref{Dinq} holds for all $(x, y) \in \mathcal{X}$  and the design given by \eqref{doptspez} is $D$-optimal. \hfill $\Box$

\subsection{Proof of Theorem \ref{thme2} }\label{sec52}
\textbf{(1)} We have to solve the $e_2$-optimal design problem in a linear regression model with a vector of regression functions given by \eqref{mat3}. For its solution  we use Elfving's theorem [see \cite{elfving1952}], which states that a design $\tilde \xi$ with masses $p_1,\ldots,p_k$ at the points $(x_1,y_1), \ldots, (x_k,y_k)$ is $e_2$-optimal if and only if there exists a constant $\gamma > 0$ and constants $\varepsilon_1, \ldots, \varepsilon_k \in \{ -1,1 \}$, such that the vector $\gamma e_2$ has the representation
\be \label{elf1}
\gamma e_2  =
\sum^k_{j=1} \varepsilon_j p_j f(x_j,y_j),
\ee
and is a boundary point of the \emph{Elving set}
\be \label{elf2}
\mathcal{R} = \mbox{conv} \big( \big \{  \varepsilon f(x,y) \mid (x,y) \in \mathcal{X} \ ; \quad \varepsilon \in \{ -1,1 \} \big \} \big)
\ee
(here conv$(\mathcal{A})$ denotes the convex hull of a set $\mathcal{A} \subset \mathbb{R}^3$). We now consider the design $\tilde\xi$ given by \eqref{xitilde1}.
 A similar reasoning as in Section \ref{sec32} shows that in the case of $j=2$ we may assume without loss of generality that $x_{\max} = y_{\max} =1$.
 Then it is easy to see that the design $\tilde\xi$ defined in \eqref{xitilde1} satisfies \eqref{elf1} with $k=2, \varepsilon_1 =  -\varepsilon_2 = 1$,
  $(x_1,y_1)=(1,1)$,  $(x_2,y_2)=(\overline{x},1)$   and
 $$
p_1= \frac {\overline{x} }{1+\overline{x} }\ ; \qquad p_2=  \frac {1}{1+\overline{x}}\ ;  \quad \gamma = \frac{\overline{x} (1-\overline{x} )}{1+\overline{x} }.
 $$
 Therefore, it remains to prove that the corresponding vector  $\gamma  e_2 =(0,\gamma,0)^T$ is a boundary point of the Elfving set $\mathcal{R}$ defined in \eqref{elf2}. For this purpose we consider the vector
 $$
 n= (n_1,n_2,n_3)^T = \Big( 1 - \frac {1+\overline{x} }{\overline{x} (1-\overline{x} )}, \frac {1+\overline{x} }{\overline{x} (1-\overline{x} )}, 0 \Big)^T
 $$
 and show that this vector defines a supporting hyperplane to the set $\mathcal{R}$ at $\gamma e_2= (0,\gamma,0)^T$, i.e.,
 \be \label{hyp1}
   (\gamma e_2)^T n &=& 1 , \\
 \label{hyp2}
   z^Tn & \leq & 1 \qquad \mbox{for all } z \in \mathcal{R}.
 \ee
Condition \eqref{hyp1} is obviously satisfied. In order to prove \eqref{hyp2} we establish the sufficient condition $|f^T(x,y)n|\leq 1$ for all $(x,y) \in \mathcal{X}$, which can be rewritten as
 \be \label{ineq}
 - \overline{x} (1-\overline{x} ) \leq xy \big(-(\overline{x}^2  +1) + (1+\overline{x} )x \big) \leq \overline{x}  (1-\overline{x} ).
 \ee
We now distinguish the cases
 \bea
 (i) & 1 \geq x \geq \frac {1+\overline{x} ^2}{1+\overline{x} } \\
 (ii) & \frac {1+\overline{x} ^2}{1+\overline{x} } > x > x_{\min}.
 \eea
 \begin{itemize}
 \item[(i)]
Here the first inequality in  \eqref{ineq} is obviously satisfied and the second (upper)  inequality holds if it can be established for $y=1$, that is
 $$
 x^2 (1+\overline{x} ) - x(\overline{x}^2 +1) - \overline{x} (1-\overline{x} ) \leq 0.
 $$
 This inequality holds for all $x \in [\frac {(-1+\overline{x} )\overline{x} }{1+\overline{x} },1]$ which proves \eqref{ineq} in the case \emph{(i)}.
 \item[(ii)]
Here  the second inequality in \eqref{ineq} is obviously satisfied and the first (lower)  inequality holds  if it can be established for $y=1$, that is
 $$
 h(x) =
 x^2(1+\overline{x} ) - x(\overline{x}^2 +1)+\overline{x} (1-\overline{x} ) \geq 0.
 $$
The minimum of this function is attained at the point $x^*=\frac {1+ \overline x^2}{2(1+ \overline x)}$. For $\bar{x}= \sqrt{2}-1$ the corresponding minimum value is given by $h(x^*) =0$ following that $h(x) \geq 0$.
For $\bar{x}= x_{\min}> \sqrt{2}-1$ the point $x^*$ is not contained in the interval $[x_{\min}, 1] $. For $x > x^*$ the function $h$ is increasing with $h(x_{\min}) = 0$ such that  \eqref{ineq} also holds in
this  case.
\end{itemize}
 Consequently, by Elfving's theorem the design \eqref{xitilde1} is $e_2$-optimal in the linear regression model defined by the vector \eqref{mat3}.\\
\textbf {(2)}
In order to prove that the design given by \eqref{xitilde3} is $e_3$-optimal in the regression model defined by \eqref{mat3}, we define the permutation matrix
$$P = \begin{pmatrix} 1 & 0 &0 \\0 & 0 & 1 \\0 & 1 & 0 \end{pmatrix} .$$
and the information matrix
\begin{equation*}
\tilde M_P(\tilde \xi) = P \tilde M (\tilde \xi) P = \int_{\mathcal{X}} xy (1, y, x)^T xy(1, y, x) d\tilde \xi(x, y)= \int_{\mathcal{X}} f(y, x) f^T(y,x) d\tilde\xi(x, y).
\end{equation*}
Note that the permutation matrix $P$ causes the exchange of the second and third component of the vector \eqref{mat3}.
Moreover, by using Lemma 9.2.4 of \cite{harv1997} we get
\bea
c^T_3 \tilde M^-(\tilde\xi)c_3 = \bigl(\tfrac{K_{ic}}{V}\bigr)^2 e^T_3 \tilde M^-(\tilde\xi)e_3 = \bigl(\tfrac{K_{ic}}{V}\bigr)^2 e^T_2 P^T \tilde M^-(\tilde\xi) P e_2 = \bigl(\tfrac{K_{ic}}{V}\bigr)^2e^T_2 \tilde M^-_P(\tilde \xi) e_2. \eea
By the first part of Theorem \ref{thme2} it follows that $(e^T_2 \tilde M^-_P(\tilde \xi) e_2)^{-1}$ is maximised by the design $\tilde{\xi}$ given by \eqref{xitilde3}.
\hfill $\Box$

\subsection{Proof of Theorem \ref{theo:extrapc1}} \label{sec53} 
\def\theequation{5.\arabic{equation}}
\setcounter{equation}{0}


Recall the definition of the  weighted polynomial regression model \eqref{eq:helpeta} and note that in this model the vector of regression functions is given by  
\begin{equation} \label{eq:helpregvec}
f(x, q) = xg(x,q) \begin{pmatrix}1 \\Êx \end{pmatrix},
\end{equation}
where the function $g$ is defined in \eqref{eq:gfun}.
The information matrix of a design $\hat \xi $ in this model is given
by
\be \label{eq:infoklein}
\hat{M}(\hat\xi) = \int_{x_{\min}}^{x_{\max}}  {f}(x,q) {f}^T(x,q)d\hat\xi( x) = \int_{x_{\min}}^{x_{\max}}  x^2g^2(x,q) (1, x) (1, x)^T d\hat\xi(x).
\ee
If the lower bound of the design space  is given by $x_{\min}=0$, an optimal design does not contain $0$ as support point. 
 In order to prove this fact  we consider the  design 
$$\hat\xi= \begin{pmatrix} 0 & x_2 & \ldots & x_k \\Ê\omega_1 & \omega_2 & \ldots & \omega_k \end{pmatrix},$$
with mass $ \omega_1>0$ at the point $0$. Now consider the 
design 
\begin{equation}\label{eq:mu}
\hat\mu= \begin{pmatrix} x_2 & \ldots & x_k \\Ê \frac{\omega_2}{1-\omega_1} & \ldots & \frac{\omega_k}{1-\omega_1}\end{pmatrix},
\end{equation}
i.e., we remove the support point $0$ and rescale the weights for the remaining support points appropriately. Then we obtain the following 
inequality with respect to the Loewner ordering
\begin{eqnarray*}
\hat M(\hat \xi) &=& \omega_1 \hat M (\delta_0) + (1-\omega_1) \hat M (\hat \mu) =  (1-\omega_1) \hat M(\hat \mu) \leq M(\hat \mu) ,
\end{eqnarray*}
where $\delta_t$ is the Dirac measure in $t$. The last inequality is strict, whenever $ M(\hat \mu)$ is non-singular.
Consequently, every design $\hat \xi$ which contains $0$ as support point can be improved by another design $\hat\mu$ of the form \eqref{eq:mu}. 
Since all optimality criteria $\Phi$  considered in this paper  (in particular the $c$-optimality criterion) are monotonically increasing
with respect to the Loewner ordering it follows that $\Phi(\hat M (\hat \xi)) \leq \Phi(\hat M (\hat \mu))$ and we 
always find an optimal design which does not contain the point $0$ in its support. Therefore we can restrict ourselves 
to the consideration of $x_{\min} >0$ throughout this section. \\
The following auxiliary result gives a solution of the $c$-optimal design problem for the vector $c=(1, 1)^T \in \mathbb{R}^2$.
\begin{lemma} \label{auxlem}
The $c$-optimal design for the vector $c=(1, 1)\in \mathbb{R}^2$  maximising 
$$
 \big( (1,1) \hat{M}^-(\hat\xi) (1,1)^T\big)^{-1} 
 $$ 
 in the class if all designs  on the interval  $ [x_{\min}, x_{\max}] \subset (0,1) $ satisfying $(1,1)^T \in $ Range$ (\hat M (\hat \xi ) )$ 
 is given by
\begin{equation} \label{eq:helpdes}
\hat \xi^*(q) = 
\begin{pmatrix} 
\bar{x}(q) & x_{\max} \\
\omega(q) & 1-\omega(q)
\end{pmatrix}
\end{equation}  
where $\omega(q)$ is defined by  \eqref{eq:helpweight} and $\bar{x}(q) $ and $ x_{\max}$ are the extremal points  
 of the equi-oscillating polynomial $\Psi(x, q)$ introduced at the beginning of Section \ref{sec42}. 
\end{lemma}

\medskip

{\bf Proof.}
Recall  that $\{x g(x, q), x^2 g(x, q) \}$  is a Chebychev system and note that the  definition of the vector $f(x,q)$ in \eqref{eq:helpregvec}
yields  $f(1, q) = c =(1,1)^T.$ As  $x_0=1 \notin [x_{\min}, x_{\max}] $  the $c$-optimal design problem for the vector $c=(1,1)^T$ 
is in fact an optimal extrapolation design  problem for the point $x_0=1$. 
Therefore it follows  from  Theorem X.7.7 in \cite{karstu1966} that  the support points of $\hat \xi^*(q)$ are given by the extremal points  of the equi-oscillating polynomial $\Psi(x, q)$ 
and the corresponding weights $\omega_1(q), \omega_2(q)$ are given by
$$\omega_1(q) = 1-\omega_2(q) = \frac{|L_1(1,q) |}{|L_1(1,q) | + |L_2(1,q)|}~, 
$$
were $L_i(x, q) = xg(x, q) (\ell_{i1} + \ell_{i2}x) $ is the $i$th Lagrange interpolation polynomial with knots $x_1= \bar{x}(q)$ and $x_2= x_{\max}$ defined by the property 
$L_i(x_j) = \delta_{ij}$, $i=1, 2$, $j=1,2$ (here $\delta_{ij}$ ist the Kronecker symbol).  
A straightforward calculation now shows  $\omega(q)$  is given by \eqref{eq:helpweight}, completing the proof of Lemma \ref{auxlem}.  \hfill $\Box$

\bigskip

It is also well known that the equi-oscillating polynomial $\Psi(x,q) $ can be expressed in terms of the optimal design $\hat \xi^*(q)$, that is 
\be \label{equi}
{\Psi}(x,q) = \frac {(1,1) \hat M^{-1}(\hat \xi^*(q))(1,x)^Txg(x,q)}
{((1,1)\hat M^{-1}(\hat \xi^*(q)) (1,1)^T)^{1/2}}
\ee
and the condition $| \underline{\Psi}(x,q) | \leq 1 $ for all $x \in [x_{\min}, x_{\max}] $ is just an equivalent formulation of the equivalence theorem 
for $c=(1,1)^T$-optimality in the regression model  \eqref{eq:helpeta} [see  \cite{karstu1966}].

\bigskip

We are now able to prove Theorem \ref{theo:extrapc1} using 
 the equivalence theorem for $c$-optimality [see \cite{pukelsheim2006}], which states that the design 
 $\tilde \xi(q^*)$  (here $q^*$ is defined by \eqref{eq:qstar}) is $c_1$-optimal  in the linear regression model defined by the vector \eqref{mat3} 
 if and only if
there exists a generalised inverse $G$ of the matrix  $\tilde{M}(\tilde\xi(q^*))$ such that the inequality
\be \label{eq:aeqc}
\left(c^T_1 G f(x, y)\right)^2 \leq \kappa := c^T_1 \tilde M^-(\tilde\xi(q^*)) c_1
\ee
holds for all   $(x,y) \in \mathcal{X}$.  
For a proof of  this statement we  
 use the  notation
$$\tilde \xi= \tilde \xi(q^*), \qquad \hat\xi^*= \hat\xi^*(q^*), \qquad g(x) = g(x, q^*), \qquad \bar{x} = \bar{x}(q^*), $$
and 
first construct an appropriate generalised inverse $G$  of the matrix $\tilde{M}(\tilde{\xi})$
which can be used in  \eqref{eq:aeqc}. 
For this purpose define 
\be \label{eq:ginv1}
G= P^{-T} H P^{-1}~, 
\ee
where the $(3\times 3)$-matrices $P$ and $H$ are given  by
\be \label{eq:ginv2}
P = \begin{pmatrix}
1 & 0 & 0 \\
0 & 1 & 0 \\
1-q^* & q^* & 1
\end{pmatrix} ,
\quad
H =  \begin{pmatrix}
\hat{M}^{-1}(\hat\xi^*) & (\frac{\sqrt{\kappa}}{\bar{x}g^2(\bar{x})}, 0)^T\\
0_2^T& 0
\end{pmatrix},
\ee
and 
 $\hat{M}(\hat\xi)$ and $\kappa $ are  defined in \eqref{eq:infoklein} and  \eqref{eq:aeqc}, respectively.
Note that $\tilde{M}(\tilde\xi)$ can be rewritten in terms of $P$ and $\hat{M}(\hat\xi)$, that is 
\begin{eqnarray*}
\tilde{M}(\tilde\xi) &=& \int_{{\mathcal{X}}} \tilde{f}(x, y) \tilde{f}^T(x, y)d\tilde\xi(x, y) =  \int_{{\mathcal{X}}} xy(1, x, y) xy(1, x, y)^Td\tilde\xi(x, y) \\
&=&  \int_{\hat{\mathcal{X}}} xg(x) (1, x, g(x)) xg(x)(1, x, g(x))^T d\hat\xi^*(x) \\
&=&  P \int_{\hat{\mathcal{X}}}  xg(x) (1, x, 0) xg(x) (1, x, 0 )^T d\hat\xi^*(x)P^T
=  P \begin{pmatrix}\hat{M}(\hat\xi^*) & 0 \\0^T_2 & 0  \end{pmatrix} P^T.
\end{eqnarray*}
By a straightforward calculation we have
\begin{eqnarray*}
\tilde{M}(\tilde\xi)G\tilde{M}(\tilde\xi)&=& P \begin{pmatrix}\hat{M}(\hat\xi^*) & 0 \\0^T_2 & 0  \end{pmatrix} P^T P^{-T} H P^{-1}  P \begin{pmatrix}\hat{M}(\hat\xi^*) & 0 \\0^T_2 & 0  \end{pmatrix} P^T \\
&=&  P \begin{pmatrix}\hat{M}(\hat\xi^*) & 0 \\0^T_2 & 0  \end{pmatrix}  \begin{pmatrix}
\hat{M}^{-1}(\hat\xi) & (\frac{\sqrt{\kappa}}{\bar{x}g^2(\bar{x})}, 0)^T \\ 0 & 0  \end{pmatrix} \begin{pmatrix}\hat{M}(\hat\xi^*)& 0 \\0^T_2 & 0  \end{pmatrix} P^T \\
&=&P  \begin{pmatrix}\hat{M}(\hat\xi^*) & 0 \\0^T_2 & 0  \end{pmatrix} P^T= \tilde{M}(\tilde\xi),
\end{eqnarray*}
and it follows that $G$ defined in \eqref{eq:ginv1} is a generalised inverse of the matrix $\tilde{M}(\tilde\xi)$.\\
A simple calculation now  shows  $\kappa =  c^T_1 \tilde M^-(\tilde\xi) c_1= (1,1)\hat M^{-1}(\hat\xi^*) (1,1)^T$ and 
using  \eqref{equi}   we obtain for the left-hand    side of \eqref{eq:aeqc} 
\begin{eqnarray*}
\left(c^T_1 G f(x, y)\right)^2 &=& \left((1, 1)  \hat{M}^{-1}(\hat\xi^*) (1, x)^T xy + \frac{\sqrt{\kappa}}{\bar{x}g^2(\bar{x})}\left(y-g(x)\right) xy \right)^2 \\
&=& \left(\frac{\sqrt{\kappa}}{\sqrt{\kappa}}(1, 1)  \hat{M}^{-1}(\hat\xi^*)(1, x)^T x g(x) \frac{y}{g(x)} +\sqrt{\kappa}\frac{g^2(x)}{g^2(\bar{x})}\left(\frac{y}{g(x)}-1\right) \frac{x}{\bar{x}}\frac{y}{g(x)}\right)^2 \\
&=& \kappa \left({\Psi}(x,q^*) \frac{y}{g(x)} + \frac{g^2(x)}{g^2(\bar{x})}\left(\frac{y}{g(x)}-1\right) \frac{x}{\bar{x}}\frac{y}{g(x)}\right)^2 \\
&:=& \kappa \tau^2(x, y)~, 
\end{eqnarray*}
where the last inequality defines the function $\tau$ in an obvious manner. \
Consequently, proving that \eqref{eq:aeqc} is satisfied for all $(x, y) \in {\mathcal{X}}$ is equivalent to proving that $|\tau(x,y)| \leq 1 $ for all $(x, y) \in {\mathcal{X}}$.\\
For the sake of transparency we therefore restrict ourselves to the case $y_{\max}=1$ and $x_{\min} \leq x_{\max}(\sqrt{2}-1)$. The other cases  can be proved in a similar way but with substantially more involved calculations.
In this case it follows that $q^*=0$ and the function $g$ reduces to $g(x,0)\equiv1$. Then the design $\hat \xi^*$ is given by
\be \label{eq:designklein}
\hat\xi^*=\hat\xi^*(0)= \begin{pmatrix}  \bar{x}  & x_{\max} \\
\frac{x_{\max}(1-x_{\max})}{x_{\max}(1-x_{\max}) + \bar{x} (1-\bar{x})} &  \frac{ \bar{x} (1-\bar{x})}{x_{\max}(1-x_{\max}) + \bar{x} (1-\bar{x}) } \end{pmatrix}, \qquad \bar{x}= x_{\max}(\sqrt{2}-1) 
\ee
with the corresponding oscillating polynomial
$
{\Psi}(x,0)= -\left(\frac{x}{\bar{x}}\right)^2 + 2\frac{x}{\bar{x}}.
$
Consequently,
\begin{equation}\label{eq:tauspecial}
\tau(x, y) =  y\frac{x}{\bar{x}}\left(y+\frac{x}{\bar{x}} -3\right).
\end{equation}
The non-zero stationary point of $\tau$ is given by $(x^*, y^*) = (\bar{x},1)$ which is a minimum of $\tau$ with $\tau(x^*,y^*) = -1$. \\
Now we consider $\tau$ on the boundary of ${\mathcal{X}}=[x_{\min}, x_{\max}]\times[y_{\min}, 1]$. First, we set $y=y_{\max}=1$ so that $\tau(x,y)$ reduces to
$$\tau_{1}(x)= \tau (x,y_{\max})= \frac{x}{\bar{x}}\left(\frac{x}{\bar{x}} -2\right),$$
which is a parabola with minimum in $x^*= \bar{x}$ and $|\tau_1(x^*)| = 1 \leq 1$. Moreover, 
$\tau_1(x_{\max})= (\sqrt{2}-1)(\sqrt{2}+1)= 1 $ and for $\tau_1(x_{\min}) \in (\tau_1(\bar{x}), \tau_1(0))= (-1, 0)$, which yields $|\tau_1(x)| \leq 1 $ for all $x\in[x_{\min}, x_{\max}]$.
Next, we set $y=y_{\min}$ so that $\tau(x, y)$ reduces to
$$\tau_{2} (x)= \tau (x,y_{\min}) =  y_{\min} \frac{x}{\bar{x}}\left(y_{\min} +\frac{x}{\bar{x}} -3\right)$$
which is again a parabola with minimum in $x^*= \tfrac{3-y_{\min}}{2}\bar{x}$ and
$|\tau_{x}(x^*) |= |\left( \tfrac{3-y_{\min}}{2}\right)^2 y_{\min}|$. Maximising the later term with respect to $y_{\min}\in (0, y_{\max})$ it follows that  $|\tau_{2}(x^*) | \leq \frac{49}{54}\leq 1$. By a similar argument it also follows that  $|\tau_{2}(x_{\max})| \leq 1$ and $|\tau_{2}(x_{\min})| \leq 1$.
Setting $x=x_{\min}$ and $x=x_{\max}$ and  using similar arguments we get $ |\tau(x_{\min},y)| \leq 1$ and  $ |\tau(x_{\max},y)| \leq 1$ for all $y\in [y_{\min}, 1]$, respectively. \\
Consequently, the inequality  $|\tau(x, y)| \leq 1$ holds for all $(x,y) \in {\mathcal{X}}$ and therefore the inequality  
\eqref{eq:aeqc} is satisfied $(x,y) \in \mathcal{X}$, which proves  the $c_1$-optimality of $\tilde{\xi}=\tilde{\xi}(0)$.
 \hfill $\Box$

\end{document}